\documentclass[aps,pre,twocolumn,nofootinbib,superscriptaddress,showpacs,preprintnumbers,notitlepage]{revtex4-1}
\usepackage{amsmath,amsfonts,amssymb}
\usepackage[utf8]{inputenc}
\usepackage{braket}
\usepackage{graphicx}
\usepackage{bm}
\usepackage{xfrac}
\usepackage{float}
\usepackage{xspace}

\usepackage{xfrac}

\DeclareMathOperator{\EX}{\mathbb{E}}
\DeclareMathOperator{\PR}{\mathbb{P}}

\newcommand{\norm}[2]{\lVert#1\rVert_{#2}}
\renewcommand{\vec}[1]{\ensuremath{\mathbf{#1}}}
\newcommand{\discriminator}{\ensuremath{C}}

\newcommand{\SM}{Supplementary Material\xspace}

\begin{document}

\title{Optical lattice experiments at unobserved conditions and scales through generative adversarial deep learning}

\author{Corneel Casert}
\email{corneel.casert@ugent.be}
\affiliation{Department of Physics and Astronomy, Ghent University, 9000 Ghent, Belgium}

\author{Kyle Mills}
\affiliation{Vector Institute for Artificial Intelligence,
  Toronto, Ontario, Canada}
\affiliation{Ontario Tech University,
  Oshawa, Ontario, Canada}
  
\author{Tom Vieijra}
\affiliation{Department of Physics and Astronomy, Ghent University, 9000 Ghent, Belgium}

\author{Jan Ryckebusch}
\affiliation{Department of Physics and Astronomy, Ghent University, 9000 Ghent, Belgium}

\author{Isaac Tamblyn}
\email{isaac.tamblyn@nrc.ca}
\affiliation{Vector Institute for Artificial Intelligence,
  Toronto, Ontario, Canada}
\affiliation{Ontario Tech University,
  Oshawa, Ontario, Canada}
\affiliation{National Research Council Canada,
   Ottawa, Ontario, Canada}

\begin{abstract}

Machine learning provides a novel avenue for the study of experimental realizations of many-body systems, and has recently been proven successful in analyzing properties of experimental data of ultracold quantum gases.
We here show that deep learning succeeds in the more challenging task of modelling such an experimental data distribution.
Our generative model (RUGAN) is able to produce snapshots of a doped two-dimensional Fermi-Hubbard model that are indistinguishable from previously reported experimental realizations.
Importantly, it is capable of accurately generating snapshots at conditions for which it did not observe any experimental data, such as at higher doping values.
On top of that, our generative model extracts relevant patterns from small-scale examples and can use these to construct new configurations at a larger size that serve as a precursor to observations at scales that are currently experimentally inaccessible.
The snapshots created by our model---which come at effectively no cost---are  extremely useful as they can be employed to quantitatively test new theoretical developments under conditions that have not been explored experimentally, parameterize phenomenological models, or train other, more data-intensive, machine learning methods.
We provide predictions for experimental observables at unobserved conditions and benchmark these against modern theoretical frameworks.
The deep learning method we develop here is broadly applicable and can be used for the efficient large-scale simulation of equilibrium and nonequilibrium physical systems.

\end{abstract}
\maketitle

Ultracold atoms provide a controlled environment for the study of emergent phenomena in many-body physical systems---including high-temperature superconductivity, many-body localization, or topological quantum phases---as well as fields such as cosmology and quantum chemistry~\cite{lewenstein_ultracold_2007, bloch_many-body_2008, gross_quantum_2017, georgescu_quantum_2014}.
Hence, finding a unifying theoretical or numerical model able to create configurations with statistics that conform to all experimental observations is of crucial importance.
Indeed, this allows one to use the model to make predictions for conditions that currently can not be experimentally realized, in addition to obtaining a better understanding of the system.
Herein we propose using generative deep learning to create such a model, that learns to produce configurations indistinguishable from those experimentally obtained and can also make predictions for configurations at larger scale or at unobserved control parameter values.
The capability of \textit{discriminative} machine learning to analyze physical systems has by now been well established, both for data obtained through numerical simulations~\cite{carleo_machine_2019, van_nieuwenburg_learning_2017, carrasquilla_machine_2017,liu_discriminative_2018,  chng_machine_2017, mills_deep_2017, mills_deep_2018}, and from experimental observations through electronic quantum matter visualization~\cite{zhang_machine_2019}, quantum gas microscopy~\cite{bohrdt_classifying_2019}, or momentum-space density images~\cite{rem_identifying_2019}.
In these machine learning applications, a neural network is trained to predict properties $\vec{y}$ of configurations $\vec{x}$, \textit{i.e.} learn the conditional probability $p(\vec{y}\vert\vec{x})$. 
Examples include the characterization of phases of matter, or efficiently calculating properties of individual microstates.
The use of \textit{generative} machine learning is relatively unexplored within experimental science; it is a much more complex problem as it requires the modeling and sampling of a probability distribution $p(\vec{x},\vec{y})$ not known \textit{a priori}. 
Yet, generative learning provides a particularly attractive approach as it relies on automatic pattern recognition---and hence does not focus on the reproduction of a specific physical quantity, which can introduce bias, or require prior knowledge about the system.
Recently, Boltzmann generators~\cite{noe_boltzmann_2019} were trained on the energy functional of many-body systems to directly generate low-energy equilibrium configurations and can overcome rare event sampling problems in simulations.
A particular class of generative models, namely restricted Boltzmann machines, has seen use as efficient variational ans\"atze for quantum many-body wave functions~\cite{carleo_solving_2017, torlai_neural-network_2018, torlai_machine-learning_2020, choo_symmetries_2018,  vieijra_restricted_nodate, melko_restricted_2019}. In return, physics-inspired algorithms (tensor networks) are now being explored for discriminative and generative tasks in machine learning~\cite{han_unsupervised_2018,stoudenmire_supervised_2016, stoudenmire_learning_2018}. \\

Here, we show that generative deep learning can be used to represent and sample the distribution of snapshots of an experimental realization of the Fermi-Hubbard model with ultracold atoms in an optical lattice. 
The format of this article is as follows: we first discuss our application area; next, we outline our generative model `RUGAN', and finally we apply it to previously reported experimental data.\\\\
\begin{center}
\begin{figure}[!ht]
            \includegraphics[width=\columnwidth]{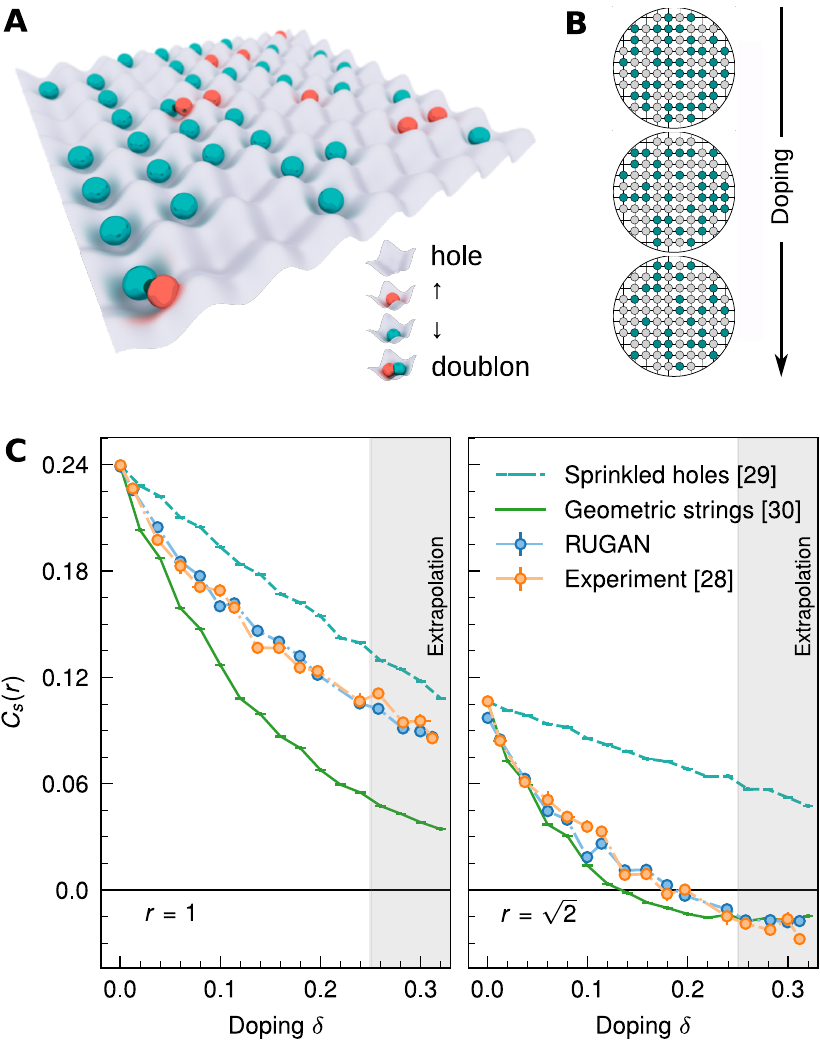}
            \caption{The Fermi-Hubbard model with ultracold atoms in an optical lattice. \textbf{A}, Experimental realization of the Fermi-Hubbard model. The two spin types  are trapped in a two-dimensional square optical lattice. The quantum state can be imaged with quantum gas microscopy. \textbf{B}, Our generative model is trained on site-resolved snapshots of quantum states, obtained at a fixed temperature and for a range of hole doping values $\delta$. Here, only one spin species is studied, while the other spin species, holes, and doublons, are all detected as empty sites. \textbf{C},
            Sign-corrected spin correlations $C_s(r)$ of Eq.~(\ref{eq:corrs}) as a function of the hole doping $\delta$ for nearest neighbors (left, $r =1$) and next-nearest neighbors (right, $r=\sqrt{2}$). The correlations obtained with our generative model `RUGAN' (1000 snapshots generated for each doping) are consistent with experiment. Note that our generative model has not been optimized on data corresponding to the largest doping values $\delta \gtrsim 0.24$ (shaded area), but is still able to produce configurations with correlations that match well with experimental observations. The geometric string theory consistently underestimates the nearest-neighbour correlations $C_s(1)$ while for both distances considered, the spin correlations obtained with sprinkled holes are overestimated. 
            }
            \label{fig:hubbard}
\end{figure}
    \end{center}
The Fermi-Hubbard model is of particular interest as it is suggested to hold the key to understanding high-temperature superconductivity \cite{lee_doping_2006, keimer_quantum_2015, gross_quantum_2017}. 
The Fermi-Hubbard model is described by the Hamiltonian 
\begin{equation}
    \widehat{\mathcal{H}} = -t \sum_{\left<\vec{i},\vec{j}\right>, \sigma}\left(\hat{c}_{\vec{i},\sigma}^\dagger\hat{c}_{\vec{j},\sigma}^{\phantom{\dagger}} + \text{h.c.}\right) + U \sum_\vec{i} \hat{c}_{\vec{i},\uparrow}^\dagger\hat{c}_{\vec{i},\uparrow}^{\phantom{\dagger}}\hat{c}_{\vec{i},\downarrow}^\dagger\hat{c}_{\vec{i},\downarrow}^{\phantom{\dagger}},
    \label{eq:hubbard}
\end{equation}
where $\hat{c}_{\vec{i},\sigma}^\dagger(\hat{c}_{\vec{i},\sigma}^{\phantom{\dagger}})$ is the creation (annihilation) operator of a spin $\sigma \in \{\uparrow, \downarrow\}$ on site $\vec{i}$.
The first term corresponds to tunneling between neighboring lattice sites $\vec{i}$ and $\vec{j}$.
The second term accounts for the on-site interaction between fermions with opposite spin.
Here we consider the strongly correlated regime, where $U/t \gg 1$.
The Fermi-Hubbard model can be experimentally realized with ultracold atoms trapped in an optical lattice (Fig.~\ref{fig:hubbard}A) \cite{lewenstein_ultracold_2007, bloch_many-body_2008, esslinger_fermi-hubbard_2010, gross_quantum_2017}.
Quantum gas microscopy provides us with site-resolved snapshots of these quantum states, imaging either the total atom distribution or that of a single spin species. 
Holes and doubly-occupied sites (doublons) are detected as empty sites (Fig.~\ref{fig:hubbard}B).
Recently, the use of discriminative machine learning for classifying snapshots of ultracold atomic gases has been investigated \cite{bohrdt_classifying_2019, rem_identifying_2019}
and we use the same data set~\cite{chiu_data_2019} as in Ref.~\cite{bohrdt_classifying_2019} to train our generative model.
At half filling, the Fermi-Hubbard model is theoretically relatively well understood and maps to the Heisenberg model with superexchange coupling \mbox{$J = 4t^2/U$}. 
In this case, long-range anti-ferromagnetic (AFM) correlations are present throughout finite-size systems for temperatures $T \ll J$. 
The Fermi-Hubbard model is not as well understood when straying from the half-filling regime through the addition of holes.
The motion of these holes displaces strings of spins and hence hides the AFM order observed at half filling; this has recently been experimentally observed~\cite{chiu_string_2019}.
These observations can be partially explained in the framework of geometric string theory, in which strings of displaced spins are added to a background of experimental snapshots produced at half filling~\cite{grusdt_parton_2018} (see \SM). 
The length of these strings is dependent on the ratio ${t}/{J}$ and the strength of the AFM correlations present in the states obtained at half filling.
Note that considering the motion of the holes is crucial to describe experimentally observed hidden order, as this is not correctly accounted for by randomly adding holes to a state obtained at half filling (``sprinkled holes").\\
The objective of this work is to develop and train a generative model capable of representing and sampling the distribution of the experimental snapshots at requested doping values. 
Such a model allows for efficient augmentation of the experimental data set, and can offer predictions for properties of microstates at dopings for which no experimental data is available, or at larger spatial scales, and hence can be used for quantitative testing of theoretical frameworks.
Our model is validated by comparing observables obtained experimentally and with the deep learning algorithm.
We first consider the AFM correlations~\cite{parsons_site-resolved_2016, hilker_revealing_2017, mazurenko_cold-atom_2017, chiu_string_2019} in snapshots created by our model---discussed in detail below---by evaluating the sign-corrected spin correlation for sites at relative displacement $\vec{r}$ with $r = \norm{\vec{r}}{2}$ and $\norm{\cdot}{p}$ denoting the $p$-norm:
\begin{equation}
    C_s(r) = 4(-1)^{\lVert\vec{r}\rVert_1}\left(\left<\hat{S}_\vec{i}^z\hat{S}^z_{\vec{i}+\vec{r}}\right>- \left<\hat{S}_\vec{i}^z\right>\left<\hat{S}^z_{\vec{i}+\vec{r}}\right> \right).
    \label{eq:corrs}
\end{equation}
More details on the calculation of these correlations are given in the \SM.
The correlations between nearest neighbours ($r=1$) and next-nearest neighbours ($r=\sqrt{2})$ measured on snapshots created by our generative model are shown in Fig.~\ref{fig:hubbard}C along with the theoretical predictions by both the geometric string theory~\cite{grusdt_parton_2018} and sprinkled holes~\cite{chiu_string_2019}. Note that neither of these theoretical models create samples with the correct correlations at large hole-doping values $\delta$ for both $r=1$ and $r=\sqrt{2}$. Our generative model is able to accurately capture the correlations across all hole-doping values---even at doping values for which it is not optimized.\\

For this work, we developed a new generative approach called a ``regressive upscaling generative adversarial network'' (RUGAN). After being shown a data set of small-scale configurations---called the training set---of a physical system, the RUGAN allows for the direct generation of new microstates at any given scale and with desired properties.
The RUGAN is able to generalize in two ways: \textit{1)} it can create microstates with properties for which no training data is available, and \textit{2)} it is able to create samples at a much larger scale (or `upscale') than the training examples.
The former is relevant for systems where obtaining configurations is numerically or experimentally feasible only for a limited set of system properties. 
As only small-scale samples are required for training, the latter generalization enables efficient sampling of configurations at scales inaccessible to traditional methods, either due to excessive computational cost or experimental restrictions on the imageable system size.\\

The RUGAN is based on generative adversarial networks (GANs) \cite{goodfellow_generative_2014, mirza_conditional_2014,arjovsky_wasserstein_2017}, which are the combination of two neural networks competing against each other as adversaries.  
One network, $G$, acts as a generator, taking samples $\vec{z}$ randomly drawn from a latent space as input and transforming these to create new samples with distribution $\PR_g \sim G(\vec{z})$. 
The other network works as a critic, learning to discern between samples coming from the generator and the example data set with distribution $\PR_r$.
These networks depend on many free parameters that are trained simultaneously, with the generator trying to trick the critic, and the critic learning how to better tell apart the generator's propositions from the training examples by measuring the distance between $\PR_r$ and $\PR_g$.
A successfully trained GAN converges to a state where the generator is so good at producing samples that the critic cannot tell the generated samples from the reference training set. 
The generator and critic of a GAN can be conditioned on additional information such as known properties of individual samples~\cite{mirza_conditional_2014}. 
This allows one to control the region of configuration space from which the generator produces new configurations.
GANs are typically used in image processing such as super-resolution \cite{ledig_photo-realistic_2016} or cross-domain pairings (\textit{e.g.}, pairing shoes with matching handbags) \cite{kim_learning_2017}, but have also been applied to the Ising model \cite{liu_simulating_2017,mills_phase_2017}, scalar field theories~\cite{urban_reducing_2018, zhou_regressive_2019}, and inverse molecular design~\cite{sanchez-lengeling_inverse_2018}.

To allow the RUGAN to create samples of arbitrary size, our generator is designed such that it consists solely of translationally equivariant operations that can be applied to latent inputs of any size.
Motivated by the locality of the interactions in the models studied here, the RUGAN consists of deep residual convolutional networks.
Convolutional neural networks, by construction, have a limited receptive field defined by the size of the convolutional kernels and the depth of the network, and can not account for arbitrary-range interactions. 
Hence, the network depth required to accurately model a physical data set gives us a proxy for the typical correlation lengths present in the individual configurations.
However, long-range interactions could also be efficiently included by making use of attention layers~\cite{zhang_self-attention_2019}.
Once optimized on the training data, such a generator can then efficiently create configurations containing a much larger number of sites.
The validity of the upscaling procedure is dependent on the same physical length scales being present in both the small-size training examples and the sizes to which we scale.
This also means that the failure of creating configurations at a larger scale ($\textit{i.e.}$, resulting in different statistics than computationally or experimentally obtained) will point us towards the appearance of physics at a new, larger length scale --- a typical example is the divergence of the correlation length at a critical point which cannot easily be captured by RUGAN.
Another advantage of the fully-convolutional design of the RUGAN is that it provides an optimized starting point for the training of larger systems.\\

Along with technical details on the design and training of the RUGAN, we give a simple illustration and results on classical spin models in the \SM.\\

\begin{figure*}[t]
            \includegraphics[width=\textwidth]{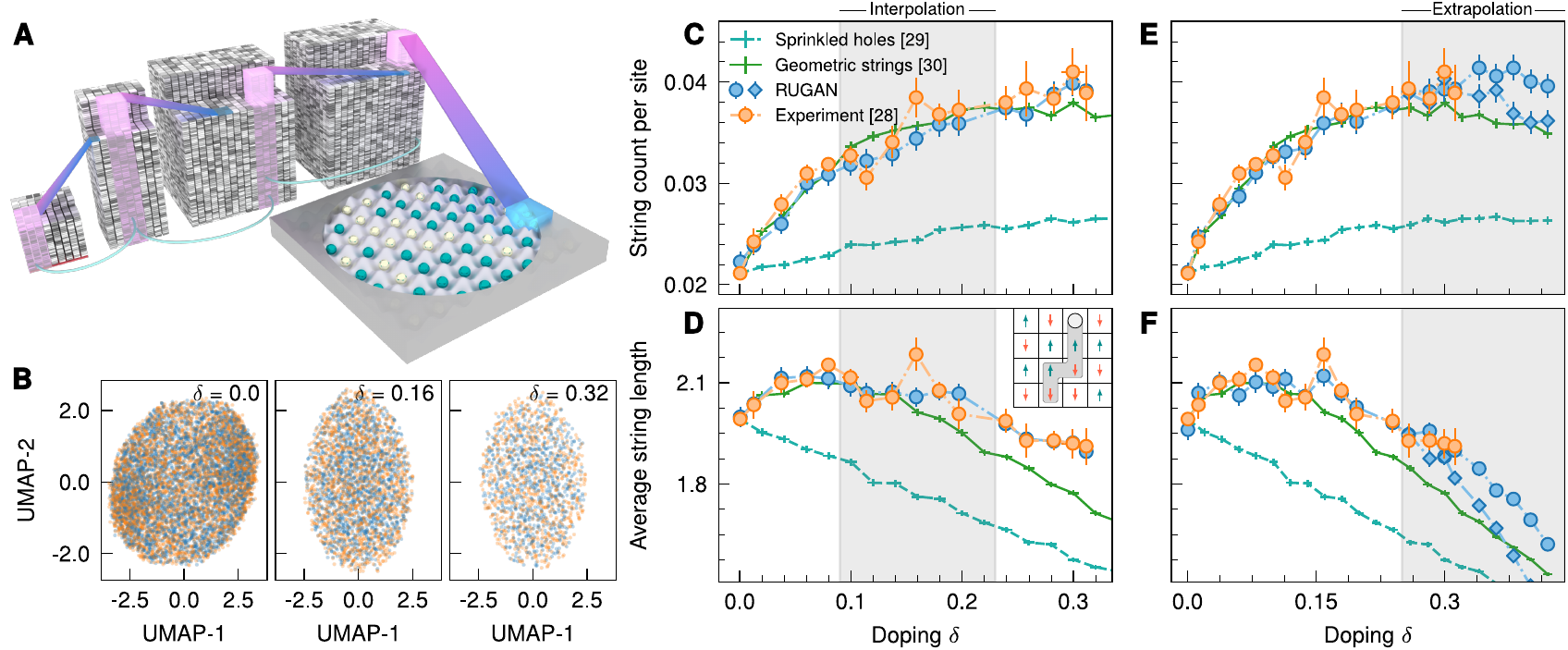}
            \caption{ String patterns in the doped Fermi-Hubbard model. \textbf{A}, Through a series of residual convolutional layers, the generative neural network transforms a latent sample to a new snapshot of a single spin species at a prescribed value of the doping $\delta$.
            \textbf{B}, The distribution of snapshots created by the RUGAN (blue) cannot be distinguished from the experimental snapshots (orange) with other unsupervised machine learning methods.
            Here, we show a dimensionality reduction with the UMAP algorithm~\cite{mcinnes_umap:_2018} of an equal number of synthetic and experimental snapshots.
            \textbf{C}, The number of strings exceeding a length of 2 per system site and \textbf{D}, the average length (sites) of the string patterns detected by the algorithm described in~\cite{chiu_string_2019}, as a function of the doping $\delta$.
            The string statistics obtained with the RUGAN (with which 1000 snapshots were created for each doping value) are consistent with experimental observations. The intermediate doping values $0.09 \lesssim \delta \lesssim 0.23$  in the shaded area are not included during training of the generative model, and the RUGAN interpolates between its knowledge at low and high $\delta$ for its creation of snapshots at these dopings. Note that both theoretical frameworks match perfectly with the experimental data at half filling by construction. For high values of the doping $\delta$, the theoretical models underestimate the average string length. (Inset) Illustration of the string pattern formation due to a hole moving through an AFM ordered state, which leaves behind a trail of displaced spins. \textbf{E} and \textbf{F}, Same as in \textbf{C} and \textbf{D}, but now the RUGAN is not provided with data corresponding to large doping values $\delta \gtrsim 0.24$ during training and is required to extrapolate to new values of $\delta$. For the extrapolation regime, we show the statistics obtained by two independently trained RUGANs. }
            \label{fig:strings}
\end{figure*}

\begin{figure}[t]
            \includegraphics{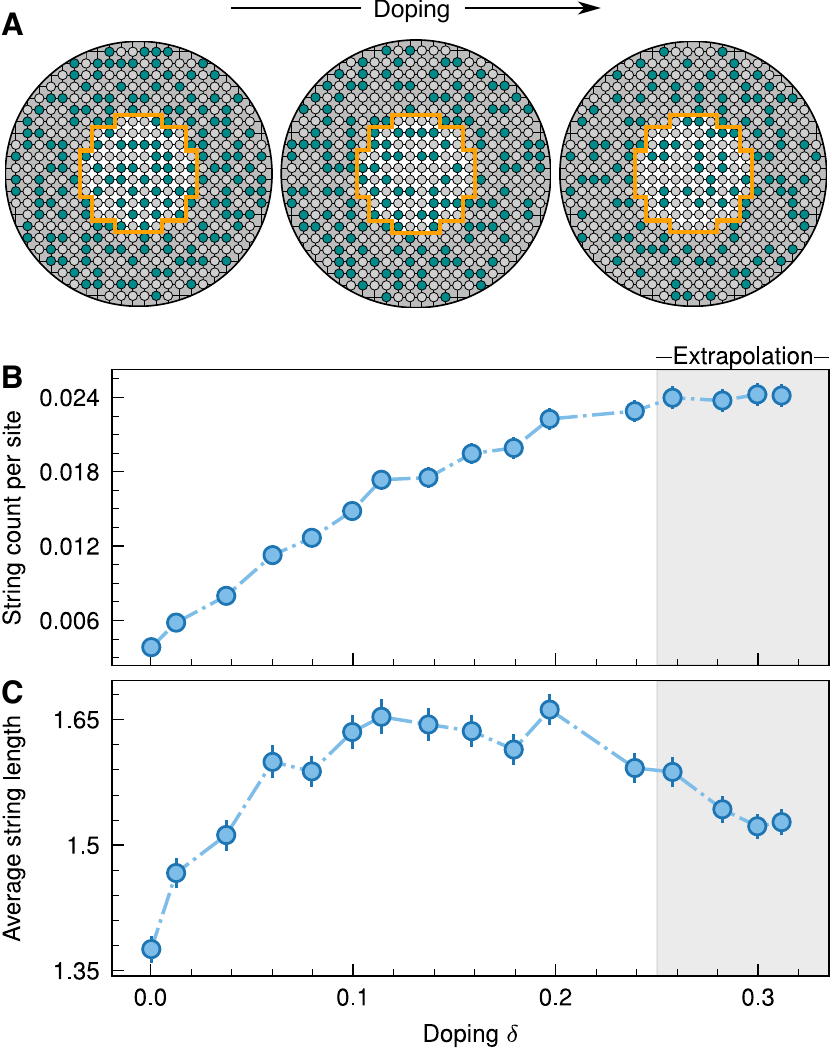}
            \caption{Size extrapolation: generating large-scale snapshots of the Fermi-Hubbard model in an optical lattice. \textbf{A}, Example snapshots of a doped Fermi-Hubbard model created at a large scale, here consisting of approximately four times as many sites as the training examples (orange line)  in Fig.~\ref{fig:hubbard}B. \textbf{B}, The number of strings exceeding a length of 2 per system site and \textbf{C}, the average length (sites) of the strings, for large-scale snapshots created with the RUGAN. For each value of $\delta$, we use the RUGAN of Fig.~\ref{fig:strings}E,F to create 1000 large-scale snapshots.}
            \label{fig:upscale}
\end{figure}

We now return to the use of generative deep learning to tackle the challenging task of modelling an experimentally obtained sample distribution. 
To this aim, we train a RUGAN on experimental snapshots of a  doped two-dimensional Fermi-Hubbard model on a square lattice, and condition it on the doping ratio $\delta$.
The output of the RUGAN is hence a series of synthetic snapshots at prescribed doping values (Fig.~\ref{fig:strings}A).
We then apply the same analysis procedure to these as to the experimentally obtained snapshots~\cite{chiu_string_2019}.
In Fig.~\ref{fig:hubbard}C, we demonstrate that a RUGAN can in this way produce synthetic snapshots with high AFM spin correlations $C_s(r)$ at small hole doping values $\delta$, and that it correctly models the decay of $C_s(r)$ in the snapshots upon increasing $\delta$.
We now show that its synthetic configurations also capture the more intricate hidden order present in the experimental snapshots, quantified by the number and average length of strings of spins displaced by hole motion as a function of $\delta$.
More information on the determination of these observables can be found in the \SM.
These statistics are shown in Fig.~\ref{fig:strings} for the synthetic samples, along with the experimentally determined values, and the predictions made by the theoretical frameworks developed in Refs.~\cite{grusdt_parton_2018, chiu_string_2019}.
Remarkably, the RUGAN is able to accurately generate snapshots that exhibit the correct string statistics even at values of $\delta$ on which it is not optimized.
In Fig.~\ref{fig:strings}C,D we show that when the RUGAN is trained on a subset of the experimental snapshots restricted to the extrema of experimentally available doping values, it still succeeds in generating configurations at intermediate $\delta$ with string statistics matching closely with those observed in experiment.
Exploiting this feature dramatically reduces the already small number of experimental observations required to train the RUGAN.
Excluding the largest values of $\delta$ from the training set allows us to assess the RUGAN's ability to extrapolate its learned knowledge of configurations with smaller $\delta$.
The observation in Fig.~\ref{fig:strings}E,F that the string statistics obtained with this extrapolation procedure again match closely with experimental results, showcases the RUGAN's capability to predict complex correlation patterns, and indicates that the synthetic configurations can serve as a benchmark for quantitative comparison with theoretical developments at conditions where no experimental data is available.
We stress that, though still performing better than theoretical predictions for doping values not included in training, the reliability of extrapolated predictions does of course eventually decrease as predictions are made at doping values farther and farther away from those used during training.
In Fig.~\ref{fig:strings}E,F, we show an example of the variation in extrapolated values that occurs far from training conditions. 
 \\

A current limitation in experiments with quantum gas microscopy on ultracold atoms is the limited number of sites that can be imaged, so that experimental snapshots of the optical lattices currently typically consist of less than 100 sites.
The upscaling ability of the RUGAN allows us to obtain a useful precursor of what could be observed at large scale while experimental realizations containing more lattice sites are still unavailable.
Given the success in performing this upscaling for classical spin models (see \SM), and the excellent agreement with experimental data for small-scale samples, these configurations can serve as a benchmark for future experimental observations of larger optical lattices.
In Fig.~\ref{fig:upscale}A, we show such configurations obtained by our RUGAN consisting of approximately four times as many sites as the experimental examples. 
As RUGAN enables us to synthesize a distribution of large-scale snapshots, we can use these to predict the behavior of physical observables at scales that are experimentally inaccessible.
In Fig.~\ref{fig:upscale}B,C we show the string count per site and average string length as a function of doping $\delta$, measured on snapshots created by the same RUGAN as in Fig.~\ref{fig:strings}E,F.\\

We have demonstrated the success of a RUGAN in modeling the distribution of experimental snapshots of a doped Fermi-Hubbard model.
To this aim, deep neural networks are trained on quantum gas microscopy images of ultracold atoms in an optical lattice.
RUGAN efficiently generates synthetic samples at requested doping values, indistinguishably distributed from the training data, and these are analyzed in the same way as one would treat experimental observations.
Whereas current theoretical frameworks of this model often focus on the description of a number of specified observables, such as spin-spin correlators or hidden order, the power of generative learning lies in its unbiased learning procedure.
Hence, especially at large doping values, the synthetic snapshots created by RUGAN provide a better match with experimental observations than current theoretical predictions.
On top of that, the RUGAN provides the ability to sample snapshots at experimentally unobserved doping values or at larger spatial scales, and thus opens the door for quantitative testing of new phenomenological, analytical, and numerical models on synthetic data under conditions where no experimental data is available. 
The observation that a RUGAN is able to make highly accurate predictions across the whole doping regime provides evidence for the existence of a unifying theoretical model.
Establishing such bounds is extremely useful across many physical domains including nucleation, phase transitions, and non-equilibrium growth.

\cleardoublepage
\section*{\SM}
\subsection{RUGAN}
\subsubsection{Generative adversarial networks}

\begin{figure*}
            \includegraphics[width=\textwidth]{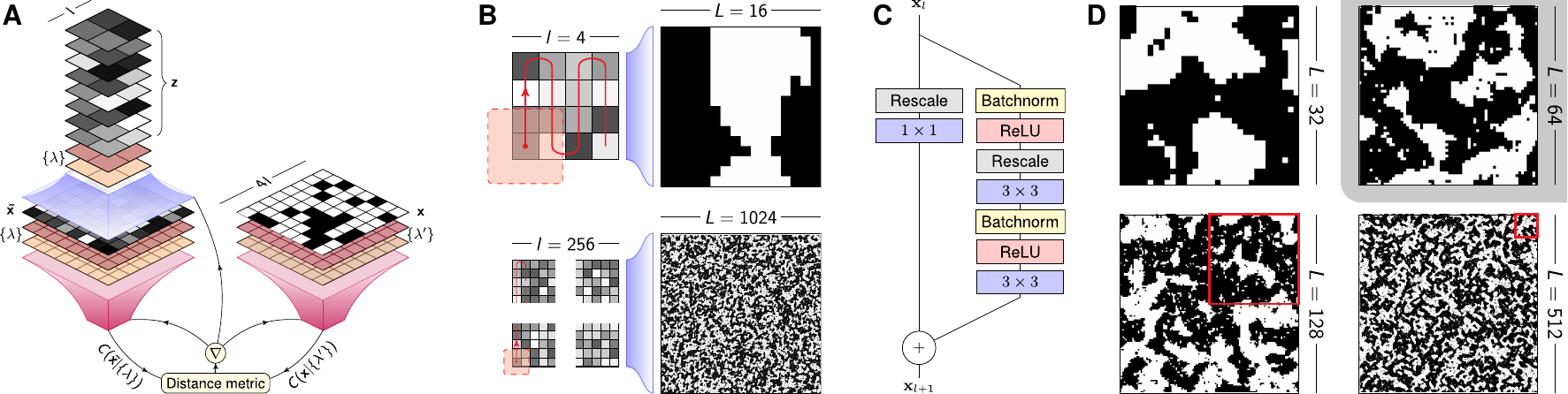}
            \caption{Generating Ising model configurations with a RUGAN. \textbf{A}, The setup of the RUGAN method. The generator $G$ (blue) transforms latent samples of size $l$, and optionally labels $\{\lambda\}$, to proposed configurations of size $4l$. The critic $C$ (pink) is used to measure the distance between the distributions of generated and real configurations. This information is used to update both neural networks. \textbf{B}, The generator consists of translationally equivariant convolutional operations and therefore can be applied to inputs of arbitrary size, allowing for the creation of configurations at different spatial scales.  
            \textbf{C}, The $l$-th hidden layer of the residual convolutional networks transforms an input $\mathbf{x}_l$ into $\mathbf{x}_{l+1}$, which can have a different spatial scale than the input. When reducing the spatial scale, the rescaling operation is an average-pooling layer with kernel size 2 and stride 2. When increasing the spatial scale, the rescaling consists of nearest-neighbor interpolation with a scale factor of 2. The convolutional operations with their corresponding kernel sizes are shown in blue. Note that we do not use batch normalization in the critic~\cite{gulrajani_improved_2017}.
            \textbf{D}, Ising configurations created by a RUGAN, trained on $L=64$ configurations, at different spatial scales. The training size is shown in red on the larger configurations.
            }
            \label{fig:architecture}
\end{figure*}

A generative adversarial network \cite{goodfellow_generative_2014} consists of a generator, which maps latent samples to new configurations, and a critic, which measures the distance between the distributions of the real samples $\PR_r$ and those of the generated samples $\PR_g$.
The distance metric used by the critic in our work is the Wasserstein-1 or Earth-Mover distance (WGAN), which allows for stable training  \cite{arjovsky_wasserstein_2017}.
The Wasserstein-1 distance $W$ between $\PR_r$ and $\PR_g$ is given by 
\begin{equation}
    W(\PR_r, \PR_g) = \inf_{\gamma \in \Pi(\PR_r, \PR_g)} \EX_{(\vec{x},\vec{y})\sim \gamma}\left[\norm{\vec{x}-\vec{y}}{2}\right],
    \label{eq:wdistance_intract}
\end{equation}
with $\Pi(\PR_r, \PR_g)$ the set of joint distributions whose marginals are $\PR_r$ and $\PR_g$.
As the infimum in Eq. (\ref{eq:wdistance_intract}) is intractable, the Kantorovich-Rubinstein duality is used to write 
\begin{equation}
    W(\PR_r, \PR_g) = \sup_{\norm{f}{L} \le 1} \EX_{\vec{x} \sim \PR_r}\left[f(\vec{x})\right] - \EX_{\tilde{\vec{x}} \sim \PR_g}\left[f(\tilde{\vec{x}})\right],
\end{equation}
with the supremum taken over all 1-Lipschitz continuous functions.
The training objective can now be formulated as a minimax game between two neural networks, the critic $\discriminator$ and generator $G$:
\begin{equation}
   \min_G \max_{\norm{\discriminator}{{L}} \le 1} \mathcal{L}(\discriminator,G),
    \label{eq:wdistance}
\end{equation} 
where 
\begin{equation}
    \mathcal{L}(\discriminator,G)= \EX_{\vec{x} \sim \PR_r}\left[\discriminator(\vec{x})\right] - \EX_{\tilde{\vec{x}} \sim \PR_g}\left[\discriminator(\tilde{\vec{x}})\right].
    \label{eq:loss}
\end{equation}
During training, the critic is optimized to maximize $\mathcal{L}$, and thus finds an estimate for $W(\PR_r, \PR_g)$. 
The generator then learns to minimize this distance, so that its sampled distribution $\PR_g$ is similar to $\PR_r$ (Fig.~\ref{fig:architecture}A).\\

The critic in the WGAN construction needs to be 1-Lipschitz continuous over the whole domain to find a correct estimate for the Wasserstein distance. 
While enforcing this constraint everywhere is impracticable, a good approximation can be obtained by adding two regularizing terms to the loss function of Eq.~(\ref{eq:loss}):
\begin{subequations}
\begin{align}
    \mathcal{L}^\prime =~&~\EX_{\vec{x} \sim \PR_r}\left[\discriminator(\vec{x})\right] - \EX_{\tilde{\vec{x}} \sim \PR_g}\left[\discriminator(\tilde{\vec{x}})\right] \\
                &+ \lambda_1  \EX_{\hat{\vec{x}} \sim \PR_{\hat{\vec{x}}}} \left[(\norm{\nabla_{\hat{\vec{x}}} \discriminator(\hat{\vec{x}})}{2}-1)^2\right]\label{eq:GP}\\
                &+ \lambda_2 \EX_{\vec{x} \sim \PR_r} \left[ \norm{\discriminator(\vec{x} + \boldsymbol{\delta}_1) - \discriminator(\vec{x} + \boldsymbol{\delta}_2)}{2}\right], \label{eq:CT}
\end{align}
\label{eq:newloss}
\end{subequations}
and as new objective 
\begin{equation}
    \min_G \max_\discriminator \mathcal{L}^\prime(\discriminator,G).
\end{equation}
A differentiable function is 1-Lipschitz continuous if it has gradients with at most unit norm over the whole domain.
Hence, the first of these regularizing terms (Eq.~(\ref{eq:GP})) penalizes the critic such that the norm of the gradient equals one for samples $\hat{\vec{x}}$ sampled from $\PR_{\hat{\vec{x}}}$ \cite{gulrajani_improved_2017}. 
As enforcing this over the whole support domain is intractable, the distribution $\PR_{\hat{\vec{x}}}$ is sampled uniformly on straight lines between data points in the training distribution $\PR_r$ and generated distribution $\PR_g$.\\
The limitation on how $\PR_{\hat{\vec{x}}}$ is sampled leaves much of the domain unconstrained.
In particular, Lipschitz continuity over the manifold that supports the training distribution $\PR_r$ is not properly enforced until the distribution $\PR_g$ lies close to $\PR_r$. 
To alleviate the lack of Lipschitz-constraint on the training manifold, a third term (Eq.~(\ref{eq:CT})) is added to the loss function that explicitly enforces Lipschitz continuity close to it \cite{wei_improving_2018}. 
Lipschitz continuity requires that for two points $\vec{x'}$ and $\vec{x''}$ close to one another, the distance $\norm{\discriminator(\vec{x'}) - \discriminator(\vec{x''})}{2}$ is bounded by a constant.
Enforcing this criterion is accomplished by perturbing every training data point twice, with small random perturbations $(\boldsymbol{\delta}_1, \boldsymbol{\delta}_2$), and minimizing the distance between the critic output of these configurations. 
In practice, the perturbation of samples is achieved by adding dropout~\cite{hinton_improving_2012}, which disables nodes in a layer with a specified probability, to several layers of the critic and feeding it the same data point twice.\\

To condition the generator on system properties, we provide it with both a random sample from the latent space and labels describing the desired properties (\textit{e.g.}, the energy, magnetization, or doping of configurations) as input~\cite{mirza_conditional_2014}.
Meanwhile, the critic is shown this same label for the generated configurations, while receiving the exact label for real samples. 
The critic uses this additional label during its estimation of the Wasserstein distance, prompting the generator to adapt by creating configurations that have features accurately described by their label.
Note that since the critic and generator find efficient internal representations of these quantities through training, they are never explicitly evaluated during training.
This implies that when we want to condition the generation on expensive operators, the vast amount of computational effort lies in generation of the small-scale samples.
Creation of new, large configurations only requires a single pass through a convolutional neural network, and hence comes at a much smaller cost.
Additionally, making generative models interpretable, \textit{i.e.} understanding how it models the interactions between the degrees of freedom~\cite{ponte_kernel_2017,wetzel_machine_2017, casert_interpretable_2019}, would lead to new insight for further theoretical developments.

\subsubsection{Neural network architectures}

Both the critic and generator are implemented as deep residual convolutional neural networks, where residual functions with respect to the layer inputs are learned (Fig.~\ref{fig:architecture}C), which allows for more stable training~\cite{he_deep_2015}. In order to be equivariant under translational operations and achieve the upscaling described in the main text, the generator consists only of convolutional layers, with no dense (i.e. fully-connected) layers that are commonly included in neural networks (Fig.~\ref{fig:architecture}C). We add a hyperbolic tangent activation function to the last layer of the generator in order to obtain a valid output representation. 
As we want to generate configurations with an approximately circular shape for the optical lattice data set, we manually apply a mask that sets the borders to zero after creating a square configuration with the generative network.\\

For the hyperparameters in our WGAN implementation, we use $\lambda_1 = 10$ and $\lambda_2 = 2$ in Eq.~(\ref{eq:newloss}) \cite{wei_improving_2018, gulrajani_improved_2017}. The dropout rate is set to 25\% for two layers in the critic to evaluate Eq.~(\ref{eq:CT}).
The weights of the neural networks are optimized with the ADAM optimizer \cite{kingma_adam:_2014}, where we set the learning rate $\alpha = 10^{-4}$ and the exponential decay rates for the first and second moment estimates to $\beta_1 = 0$ and $\beta_2 = 0.9$.
To gain a more reliable estimate of the Wasserstein distance before updating the generator's weights, on the experimental data we train the critic on 20 batches for every generator training iteration. For the classical spin models we use a 10:1 ratio of critic updates to generator updates. Each model is trained for 2000 epochs.\\

Details on the network architectures (\textit{e.g.}, number of layers and channels)  can be found in our open-source implementation at \url{http://clean.energyscience.ca/codes}.

\begin{figure}[b!]
            \includegraphics[width=\columnwidth]{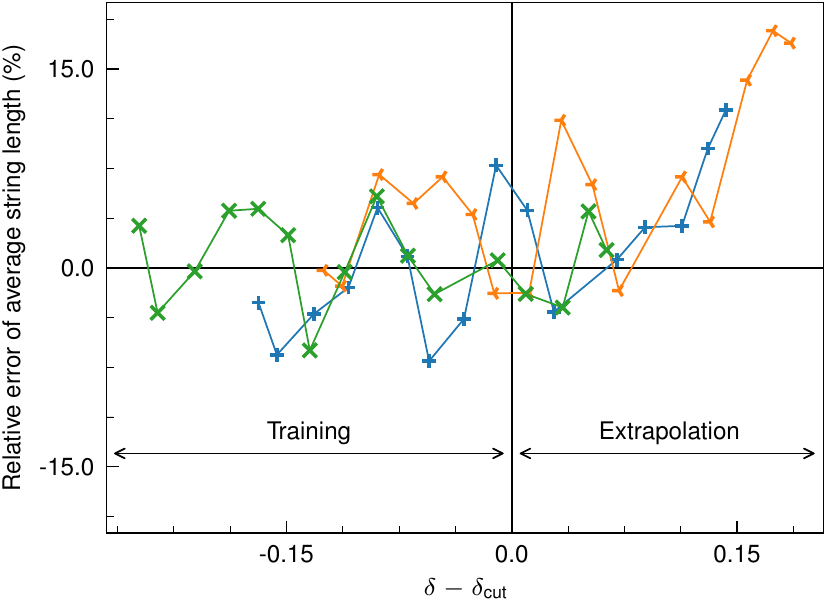}
            \caption{Relative error on the average string length for RUGANs trained on different subsets of the available data. For the green line, the doping values on which we train are cut off to only include the 12 lowest values (same as in Fig.~\ref{fig:strings}F); this is 9 and 7 for the blue and orange line respectively.
            }
            \label{fig:residuals}
\end{figure}
            
\subsubsection{Model selection}

Once training is complete, we use each model epoch to generate a number of configurations for every conditioning label. 
As the Wasserstein distance quickly stabilizes (after a couple of epochs) to a constant value during training, we resort to a different selection criterion to decide on which model is ultimately deployed.
Here, we evaluate the squared deviation between several observables (as shown in the main text) measured with the experimental and RUGAN snapshots (weighted by the experimental error) and select the model that minimizes this deviation.
Naturally, when data corresponding to certain conditioning labels is not shown during training (\textit{i.e.}, for the demonstration of interpolation and extrapolation), data generated at these labels is also not used for model selection, and we select the model that performs best on the training regime.
The results obtained when interpolating between the lowest and highest doping values (Fig.~\ref{fig:strings}C,D) consistently match well with experimental values across the model epochs.
As expected, in the case of extrapolation, the results when generating data at doping values much higher than those trained on are less robust, as detailed in Fig.~\ref{fig:strings}E,F.
We also demonstrate this in Fig.~\ref{fig:residuals}, where we provide the RUGAN with even smaller subsets of the available doping values than in Fig.~\ref{fig:strings}E,F. Though again better than theoretical predictions, the string statistics measured on the synthetic configurations start to deviate from the experimental observations for the highest doping values, even for the model epochs that perform best on the training set.

\subsubsection{An example: the Ising model}

\begin{figure}[b!]
            \includegraphics[width=\columnwidth]{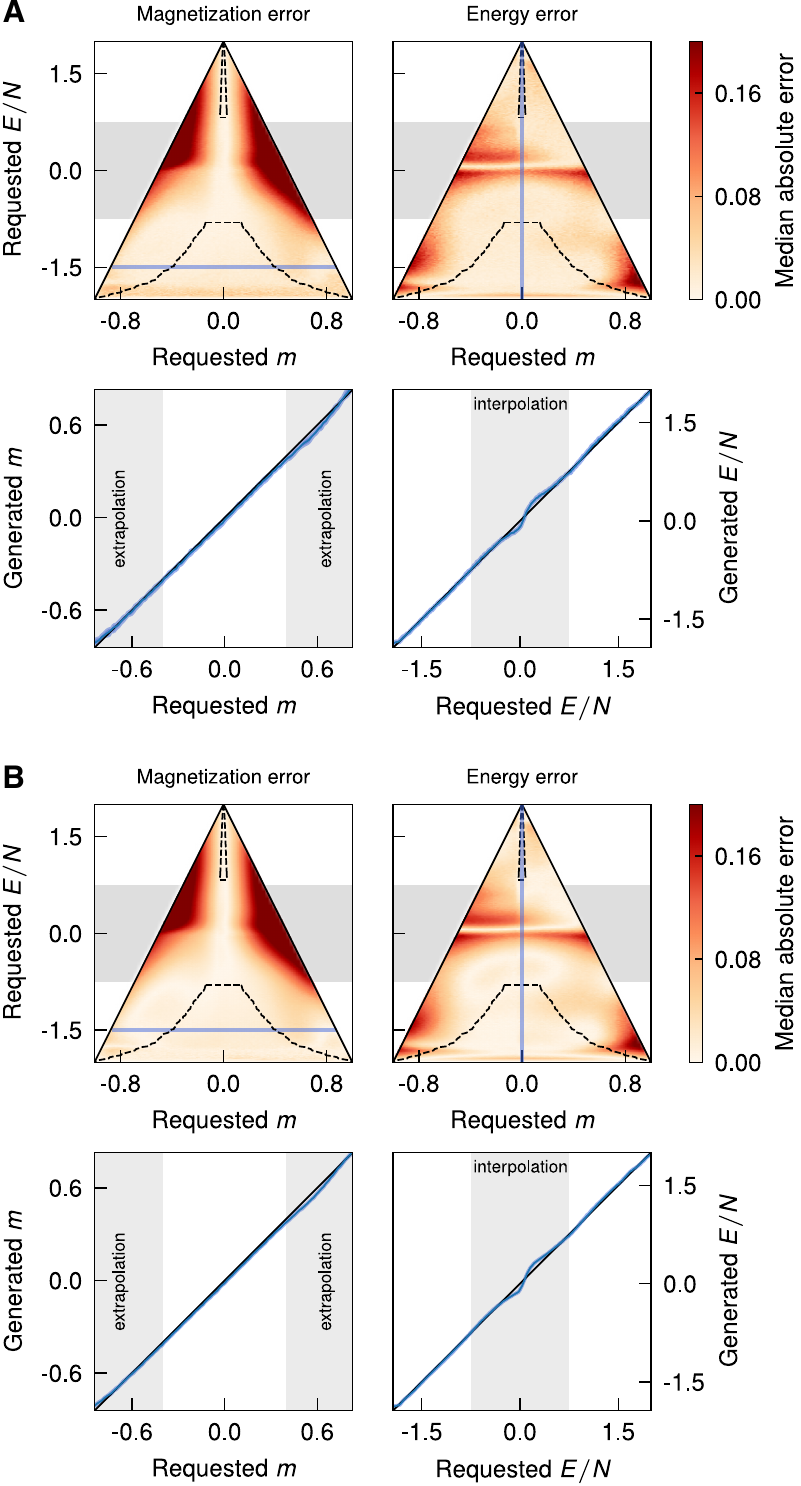}
            \caption{Generating Ising model configurations with a RUGAN. \textbf{A}, During training, we condition the RUGAN on the energy and magnetization of each configuration, and only show it examples with labels in the regions enclosed by dashed lines. After training, it is requested to create configurations with all possible labels. The median absolute error made here is shown in the top row. In the bottom row, we show the average error and its standard deviation at fixed energy (left) and fixed magnetization (right), indicated by the blue lines in the top row.
            The results are obtained by sampling 100 configurations at possible combinations of the energy and magnetization per site with energy spacing  $\Delta(E/L^2) = 1/64$ and magnetization spacing $\Delta(m) = 1/1024$. Results shown here are for the training system size $L=64$. \textbf{B}, Same as A, but now with configurations created at a larger scale $L = 256$, containing 16 times more spins than the training examples.
            }
            \label{fig:ising}
\end{figure}

We now give a detailed illustration of our framework on the prototypical classical Ising model on a two-dimensional square lattice of length $L$ with Hamiltonian ${\mathcal{H}} = -J\sum_{\left<\mathbf{i},\mathbf{j}\right>} s_\mathbf{i} s_\mathbf{j}$, where the sum runs over nearest-neighbor spins and we set $J = 1$.
For $N$ binary spins $s \in \{-1, +1\}$, the configuration space of the Ising model has a dimension of $2^{N}$; sampling configurations with desired properties directly from this space is intractable for all but trivial system sizes.
Here we show that this task can be accomplished with a RUGAN by training it on a data set of small-scale Ising configurations, and conditioning it on the energy and magnetization density $m = \sum_i s_i/N$ of each training example.
The conditioning allows for the efficient creation of microstates with desired properties from the high-dimensional configuration space as well as the creation of microstates with conditioning labels for which no training examples are available.
We implemented a modified form of umbrella sampling called ``targeted sampling''~\cite{mills_deep_2018} so that we can obtain a training set with a uniform energy distribution. This sampling method resembles the Metropolis-Hastings algorithm in structure, but instead of seeking low-energy states, we target specific energies, accepting configurations that move us toward the target energy, and rejecting ones (with a Gaussian probability) that lead us away. In doing this, we can collect examples across the energy spectrum in an efficient way.
In Fig.~\ref{fig:ising}A, we train a RUGAN on a data set of Ising microstates restricted to high and low energy values and magnetizations near zero, and use it to sample the entire space of possible energy and magnetization combinations.
The generator only makes large errors on the conditioning label combinations with a relatively small density of states.
Note that spin-flip symmetry is not enforced, which could potentially decrease the errors shown here. 
The accuracy in this conditioning is retained when using the upscaling property to create configurations at much larger scales (Fig.~\ref{fig:ising}B).
This implies that we can greatly accelerate sampling of uncorrelated large scale synthetic configurations, as costly simulations are only needed for a small subset of the configuration space and only of small-scale microstates.\\

\subsection{Hidden order in the two-dimensional Fermi-Hubbard model}

To benchmark the string patterns in the experimental data and our RUGAN, we compare them to the frameworks of sprinkled holes and geometric string theory~\cite{grusdt_parton_2018} and apply an analysis procedure identical to what has been previously used to describe experimental observations~\cite{chiu_string_2019}. For simplicity, we here recapitulate these but point towards Refs.~\cite{grusdt_parton_2018, chiu_string_2019} for more detailed information. \\

\subsubsection{Sprinkled holes and geometric string theory}

Sprinkled holes is a model for the doped Fermi-Hubbard model in the limit of non-interacting holes.  
To obtain snapshots at different dopings, we start from experimentally obtained snapshots at half filling and add holes on random positions until the doping matches the requested one.\\

Geometric string theory is a theoretical model where holes do not interact with each other but do interact with the surrounding spins.  
First, a single hole is placed at a random position on the lattice.  
The dynamics of the hole can be described by introducing an effective Hamiltonian and an effective Hilbert space for a single string~\cite{chiu_string_2019, grusdt_parton_2018}.  
The Hilbert space consists of string patterns, which can be viewed as paths without loops on a Cayley tree with coordination number $z=4$.  
Using the frozen spin approximation and $U \gg t$, the strings can be modeled by the effective Schr\"odinger equation
\begin{equation}
    t\sum_s^{z-1}{\psi_{l+1,s}} + t\psi_{l-1} + V_l\psi_l = E \psi_l,
\end{equation}
where $\psi_{l}$ is a shorthand notation for a path on the Cayley tree of length $l$, and $\psi_{l+1,s}$ denotes the string obtained by continuing the string $\psi_l$ along one of the $z-1$ directions on the Cayley tree. 
The parameter $t$ is the coupling constant for tunneling between string lengths (equal to $t$ in the Fermi-Hubbard Hamiltonian) and $V_l$ is an effective potential.  
The effective potential $V_l=(\text{d}E/\text{d}l)l + g\delta_{l,0}$ consists of a linear tension with magnitude $\text{d}E/\text{d}l$ and an attractive term with magnitude $g$.  
Solving this string model for finite temperature yields a string length distribution $p(l)$.  
The snapshots are then created by starting from experimental snapshots at half filling and adding strings on random positions in the lattice---with a length according to the string length distribution $p(l)$---until the desired doping is reached. 
More details can be found in Refs.~\cite{grusdt_parton_2018, chiu_string_2019}.

\subsubsection{String detection algorithm}
In the main text the number of strings and their average length as a function of doping are compared between the RUGAN, experiment and the theories explained above. The detection algorithm of these strings is applied to snapshots where one of the two spin species and doublons are removed, and is performed in multiple steps~\cite{chiu_string_2019}.  The geometric strings describe the deviation between the doped Fermi-Hubbard snapshots and a checkerboard state. Hence, the first step involves selecting a window (here with a diameter of 7 sites) for each configuration with the highest staggered magnetization. Using a window with a diameter smaller than the configuration itself negates some of the finite temperature effects. For a given doping, 60\% of the resulting windows with the highest staggered magnetization are kept for further analysis. In the next step, each of these windows is compared to a checkerboard state. The strings are then identified as deviations from this checkerboard, and the string-pattern length distribution $p^\delta(l)$ is measured.

As for the string count shown in Fig.~\ref{fig:strings}, only those patterns of length greater than two are included as to negate the contribution from quantum fluctuations such as doublon-hole pairs~\cite{chiu_string_2019}. The average string lengths $\bar{l}(\delta)$ are calculated from the string length histograms as \mbox{$\bar{l}(\delta) = \sum_l lp^\delta(l)/\sum_l p^\delta(l)$}.

\subsubsection{Spin-spin correlators}

One way to assert the validity of the snapshots created by the RUGAN described in the main text is to verify whether the sign-corrected two-point spin correlator \begin{equation}
    C_s(r) = 4(-1)^{\lVert\vec{r}\rVert_1}\left(\left<\hat{S}_\vec{i}^z\hat{S}^z_{\vec{i}+\vec{r}}\right>- \left<\hat{S}_\vec{i}^z\right>\left<\hat{S}^z_{\vec{i}+\vec{r}}\right> \right)
\end{equation}
    matches well with experimental results. 
    Here, \mbox{$\hat{S}_\vec{i}^z = \tfrac{1}{2} (\hat{n}^\uparrow_\vec{i} - \hat{n}^\downarrow_\vec{i})$} with $\hat{n}^\sigma_\vec{i}$ the number operator for spin $\sigma$ on site $\vec{i}$ .
The spin correlator can be calculated from the experimental snapshots as~\cite{parsons_site-resolved_2016}\begin{equation}
\begin{split}
    C_s(r) = (-1)^{\lVert\vec{r}\rVert_1}\bigg[&2\sum_{\sigma\in \{\uparrow, \downarrow\}}\left(\left<pp\right>_{R\sigma} - \left<p\right>^2_{R\sigma} \right) \\ &- \left(\left<pp\right>_{NR} - \left<p\right>^2_{NR} \right) \bigg],
\end{split}
\end{equation}
where the spatial indices are dropped for simplicity.
Here, $p$ denotes a singly occupied site, the expectation value $\left<\cdot\right>_{NR}$ is taken over images where neither spin species was removed, and $\left<\cdot\right>_{R\sigma}$ over images where the spin state $\sigma$ was removed.
To calculate the expectation values $\left<\cdot\right>_{NR}$, we train a second RUGAN on a data set of snapshots containing both spin species, and also condition it on the doping.
Here, the doping conditioning can be explicitly checked, as the doping $\delta \approx 1.22 (0.905-n_s)$ where $n_s$ is the density of singly-occupied sites~\cite{chiu_string_2019}.

\subsubsection{Data set}

The experimental data of the Fermi-Hubbard model is obtained from Ref.~\cite{chiu_data_2019}.
These data are obtained at temperature $ T = (0.65\pm0.04)J $, and the ratio $U/t = 8.1(2)$. 
Here, a mixture of the two lowest hyperfine states of $^6$Li is trapped in a two-dimensional optical lattice. 
Site-resolved measurements of the occupation in the optical lattice are obtained with high-resolution quantum gas microscopy \cite{mazurenko_cold-atom_2017}. 
The experimental snapshots of the atomic distributions consist of a circular region of 80 sites. 
In total, 8822 images of the atomic distributions with both spin components present and 17233 with one spin component removed are available.
We augment the data set by also including these samples rotated by multiples of $90^\circ$.

\subsection{Acknowledgements}
We thank Christie S. Chiu, Phil de Luna, Dennis Klug, Roger Melko, Jannes Nys and Michael Schuurman for helpful discussions.
Work at NRC was carried out under the auspices of the MCF and AI4D Program.  
The computational resources and services used in this work were partly provided by the VSC and the Flemish Government – department EWI.
T.V. is supported as an ‘FWO-aspirant’ under contract number FWO18/ASP/279.
K.M. \& I.T. acknowledge support from NSERC. 
The Titan V used for this research was donated by the NVIDIA Corporation.

\bibliography{references}

\end{document}